\documentclass[aps,showpacs,prl,twocolumn,superscriptaddress,apsrev,english]{revtex4-1}
\usepackage{multirow}
\usepackage{color}
\usepackage{amsfonts}
\usepackage{amsmath}
\usepackage{mathrsfs}
\usepackage{amssymb}
\usepackage{wasysym}
\usepackage{graphicx}
\usepackage{babel}
\usepackage{paralist}
\usepackage{indentfirst}
\usepackage{subfigure}
\usepackage{hyperref}
\usepackage{ulem}
\usepackage{soul}
\usepackage{cancel}

\begin{document}
\title{Doubly Modulated Optical Lattice Clock: Interference and Topology}
\author{Xiao-Tong Lu }
\thanks{These authors contributed equally to this work.}
\affiliation{Key Laboratory of Time and Frequency Primary Standards, National Time Service Center, Chinese Academy of Sciences, Xi'an 710600, China}
\affiliation{School of Astronomy and Space Science, University of Chinese Academy of Sciences, Beijing 100049, China}
\author{Tao Wang }
\thanks{These authors contributed equally to this work.}
\affiliation{Department of Physics, and Center of Quantum Materials and Devices, Chongqing University, Chongqing, 401331, China}
\affiliation{Chongqing Key Laboratory for Strongly Coupled Physics, Department of Physics, Chongqing, 401331, China}
\author{Ting Li}
\affiliation{Key Laboratory of Time and Frequency Primary Standards, National Time Service Center, Chinese Academy of Sciences, Xi'an 710600, China}
\affiliation{School of Astronomy and Space Science, University of Chinese Academy of Sciences, Beijing 100049, China}
\author{Chi-Hua Zhou}
\affiliation{Key Laboratory of Time and Frequency Primary Standards, National Time Service Center, Chinese Academy of Sciences, Xi'an 710600, China}
\affiliation{School of Astronomy and Space Science, University of Chinese Academy of Sciences, Beijing 100049, China}
\author{Mo-Juan Yin}
\affiliation{Key Laboratory of Time and Frequency Primary Standards, National Time Service Center, Chinese Academy of Sciences, Xi'an 710600, China}
\affiliation{School of Astronomy and Space Science, University of Chinese Academy of Sciences, Beijing 100049, China}
\author{Ye-Bing Wang}
\affiliation{Key Laboratory of Time and Frequency Primary Standards, National Time Service Center, Chinese Academy of Sciences, Xi'an 710600, China}
\affiliation{School of Astronomy and Space Science, University of Chinese Academy of Sciences, Beijing 100049, China}
\author{Xue-Feng Zhang }
\email{zhangxf@cqu.edu.cn}
\affiliation{Department of Physics, and Center of Quantum Materials and Devices, Chongqing University, Chongqing, 401331, China}
\affiliation{Chongqing Key Laboratory for Strongly Coupled Physics, Department of Physics, Chongqing, 401331, China}
\author{Hong Chang }
\email{changhong@ntsc.ac.cn}
\affiliation{Key Laboratory of Time and Frequency Primary Standards, National Time Service Center, Chinese Academy of Sciences, Xi'an 710600, China}
\affiliation{School of Astronomy and Space Science, University of Chinese Academy of Sciences, Beijing 100049, China}
\begin{abstract}
	 The quantum system under periodical modulation is the simplest path to understand the quantum non-equilibrium system, because it can be well described by the effective static Floquet Hamiltonian. Under the stroboscopic measurement, the initial phase is usually irrelevant. However, if two uncorrelated parameters are modulated, their relative phase can not be gauged out, so that the physics can be dramatically changed. Here, we simultaneously modulate the frequency of the lattice laser and the Rabi frequency in an optical lattice clock (OLC) system. Thanks to ultra-high precision and ultra-stability of OLC, the relative phase could be fine-tuned. As a smoking gun, we observed the interference between two Floquet channels. Finally, by experimentally detecting the eigen-energies, we demonstrate the relation between effective Floquet Hamiltonian and 1-D topological insulator with high winding number. Our experiment not only provides a direction for detecting the phase effect, but also paves a way in simulating quantum topological phase in OLC platform.
\end{abstract}
\maketitle

\textit{Introduction.--} Floquet engineering is a powerful tool for quantum simulating exotic Hamiltonian via time-periodic driving \cite{Eckardt_2017,Kitamura_2019,Kiefer_2019,Roushan_2016,Shu_2018,Mukherjee_2018,Mei_2020}.
Extending modulation from one-parameter to two is a tantalizing non-trivial task in both physics and technique aspects. The relative phase will strongly change physics due to the time-reversal symmetry breaking \cite{Roushan_2016,Wen_1989,Esslinger_2019}, and the typical phenomena is the observation of interference. Various theoretical proposals of double modulation are discussed, such as realizing unconventional Hubbard model \cite{Poletti_2014,Mintert_2019}, studying quantum scar \cite{Knolle_2020} and controlling particles migration \cite{Yang_2016}. However, the experimental realization is extremely hard in the quantum many body systems, because technically it requires both parameters to be independently fine-tuned to make relative phase ultra-stable, and the resulting phenomena can be clearly observed.

\begin{figure}[t!]
	\includegraphics[width=0.99\linewidth]{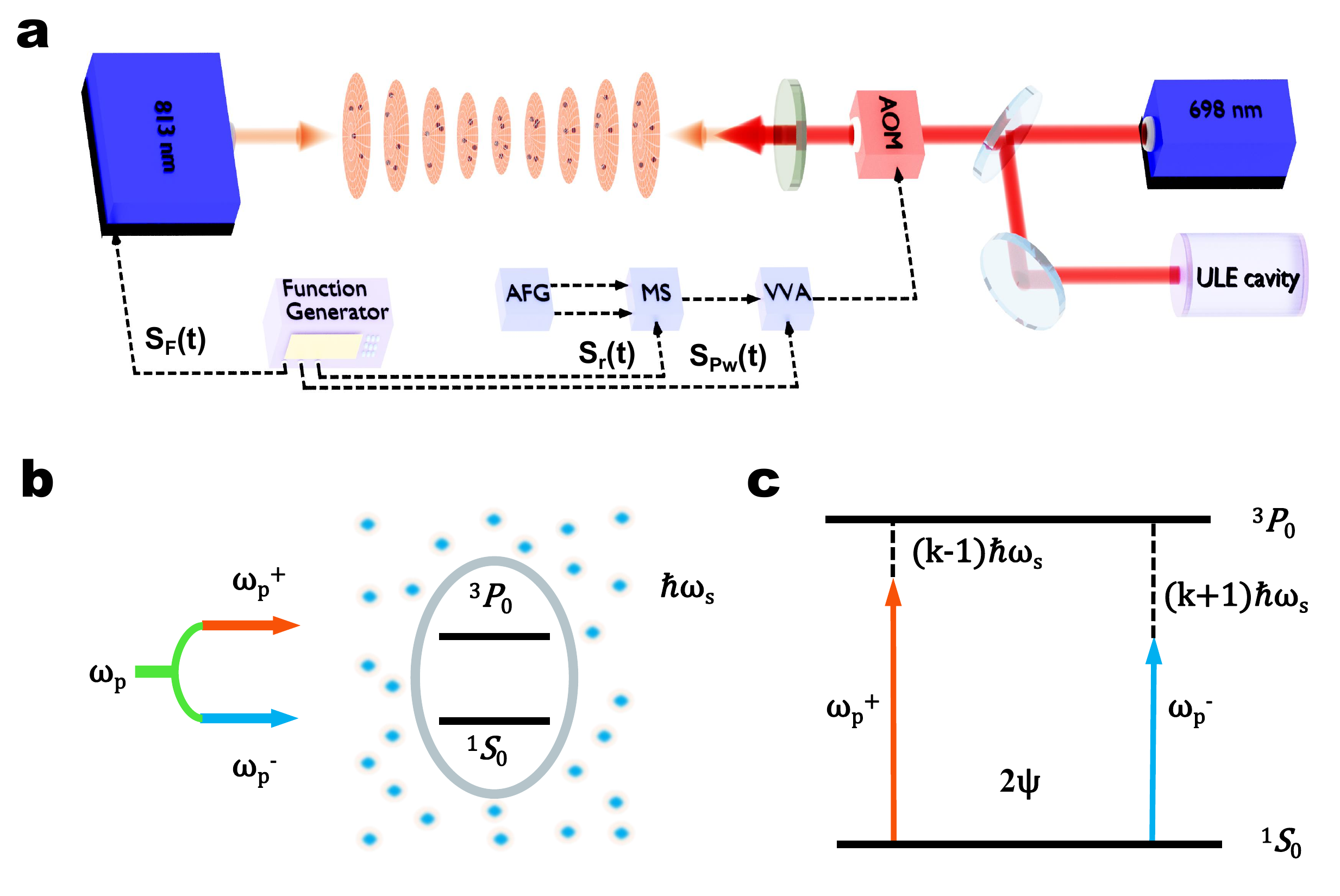}\hfill
	\caption{\label{fig1} Schematic picture of experiment setup and interference. (a) The function generator provides three signals S$_{\mathrm{F}}(t) $, S$_{\mathrm{r}}(t) $ and S$_{\mathrm{PW}}(t) $. S$_{\mathrm{F}}(t) $ and S$_{\mathrm{PW}}(t) $ are used for modulating the frequency of 813 nm lattice laser and the power of clock laser by the voltage variable attenuator (VVA), respectively.  The signal S$_{\mathrm{r}}(t) $ can be input into the microwave switch (MS) to switch two signals 0 and $\phi$ generated by the arbitrary function generator (AFG), so that the phase shift due to the transverse amplitude modulation can be compensated. The phase, power and frequency of the 698 nm clock laser are simultaneously changed by imposing the signal from VVA to the acousto-optic modulator (AOM). (b) Under the longitude modulation, the two-level atom can be treated as surrounded by `Floquet photons' with energy $\hbar \omega_s$ in dressed atom picture. (c) The transverse modulation provides two Floquet channels: $\omega_p^+$ and $\omega_p^-$ through which the atoms can hop from $^{1}$S$_0$ to $^{3}$P$_0$ with the assistant of longitude `Floquet photons'. The relative phase $2\psi$ between two hopping processes causes the interference.} 
\end{figure}
As one of the most accurate platforms, the optical lattice clock system becomes an ideal candidate \cite{Katori_2005,Ye_2015,PoliNC_2013}. The OLC is composed of an optical local oscillator stabilized by an appropriately chosen two energy levels transition of thousands of atoms trapped in a lattice potential \cite{Katori_2003,Westergaard_2011,Ye_2018,Ye_2020}. The lattice laser at `magic-wavelength' guarantees both energy levels feel the same lattice potential and the line-width of spectrum can be suppressed to several Hz. Meanwhile, the long lifetime of excited state keeps the system unaffected by spontaneous emission during measurement. Thus, it is not only taken as candidate of the next standard of time, but also a perfect platform for measurements of fundamental constants in physics \cite{Safronova_2018,kolkowitz_2016,norcia_2017}.

In this manuscript, we successfully design and implement an experiment to simultaneously modulate the internal levels of atom as well as the Rabi frequency in $^{87}$Sr OLC system (Fig.\ref{fig1} (a)) \cite{Katori_2005,Ye_2015,PoliNC_2013,Ye_2017}. It can be taken as an effective magnetic field $\vec{h}(t)=\{h_x(t),0,h_z(t)\}$ coupled to two energy levels in description of Pauli matrix $\vec{\sigma}$. The Hamiltonian can be written as 
\begin{equation}
\hat{H}=\hbar\vec{h}(t) \cdot \vec{\sigma}.\label{H}
\end{equation} 
The longitude field $h_z(t)$ is equal to $(\delta+A\cos(\omega_s^z t+\psi))/2$ where $\delta=\Delta-\omega_p$ is detuning frequency between atom internal energy gap $\Delta$ and clock laser photons $\omega_p$, and the associate modulation is characterized with longitude driving frequency $\omega_s^z$, amplitude $A$ and phase $\psi$. Meanwhile, the transverse field $h_x(t)$ is $\Omega_{\vec{n}}\cos(\omega_s^x t)/2$, where $\Omega_{\vec{n}}$ is the static Rabi frequency of external harmonic oscillator state $|\vec{n}\rangle$ without driven and $\omega_s^x$ is transverse driving frequency. In order to check the phase effect, we set $\omega_s^x=\omega_s^z=\omega_s$. Owing to the highly precision spectrum in OLC platform \cite{Ye_2009}, the Floquet sidebands can be resolved at driving frequency down to $100$Hz \cite{SM}, so that we can stabilize and fine-tune the relative phase which remarkably changes the movement path of $\vec{h}(t)$. Because of longitude modulation, the transition between two levels can be assisted by longitude `Floquet photons' with the energy $\hbar \omega_s$. On the other hand, the transverse modulation of clock laser results in two Floquet channels with `photon frequency' $\omega^-_p=\omega_p-\omega_s$ and $\omega^+_p=\omega_p+\omega_s$ respectively. As shown in Fig.\ref{fig1}(b-c), when the energy gap $\Delta$ is equal to $\omega^-_p+(n_d+1)\omega_s=\omega^+_p+(n_d-1)\omega_s$ ({$n_d$ is the order of Floquet sideband}), the initial relative phase can induce the interference between 'Floquet photon' assisted tunnelings through two different transverse Floquet channels. Furthermore, the effective Floquet Hamiltonian can be taken as two effective magnetic fields with different winding numbers coupled to atoms. It is directly related to the 1-D topological insulator, and different Floquet sideband exhibits strong relations with the high order winding number.

\textit{Experiment setup.--}Approximately $5\times10^4$ fermionic $^{87}$Sr atoms are cooled down to  $\approx3\mu$K  and loaded into a quasi one-dimensional optical lattice aligned with the $z$ axis (experimental details shown in supplementary materials \cite{SM}). As shown in Fig.\ref{fig1}(a), the lattice is created by a counter-propagating laser beam at magic-wavelength $\lambda_{L}= 813.42$ nm, so that the dipole-forbidden transition energy levels $(5s^{2})^{1}$S$_{0} (|g\rangle)$ and $(5s5p)^{3}$P$_{0}(|e\rangle)$ feel the same lattice potential. About $N_L\approx1200$ lattice sites are separated by barriers of height $V_{0}\approxeq 90E_{r}$ ($E_{r}$=$3.44$kHz the recoil energy) which is strong enough to hinder inter-site tunneling. The lifetime of excited state is about $160$s, while each round of the experiment lasts for several hundreds of milliseconds, so the spontaneous emission could be ignored. Our platform could be well described as an ensemble of non-interacting atoms taking pseudo-spin half trapped in a harmonic trap \cite{SM}. All the atoms are prepared at the ground internal states $|g\rangle$, while the distribution at external states $|\vec{n}\rangle$ follows the Boltzmann's distribution law.  Then, after interacting with a {plane-wave} clock laser at Lamb-Dicke regime \cite{SM}, the atoms follow Rabi oscillation between the internal states with different Rabi frequency according to their external states, and the bare Rabi frequency is { $\Omega_0/2\pi\approxeq 9.0$Hz}.

\textit{Longitude modulation.--}{The frequency of lattice laser is modulated as $\omega_L(t)=\bar{\omega}_L+\omega_a \sin(\omega_s t+\psi)$ with $\bar{\omega}_L=2\pi c/\lambda_{L}$ and $\omega_a$ is the driving amplitude \cite{SM,Chang_2020}.}
In the lattice co-moving frame, due to the relativistic Doppler effect, the atoms feel an effective clock frequency $\omega_p'(t)=\omega_p-A\cos(\omega_s t+\psi)$ with driving amplitude $A=\omega_a\omega_p\omega_sL/c\bar{\omega}_L$ ($L$ is the distance from the high-reflecting mirror to the lattice center). Then, the system can be approximately described with time-dependent Hamiltonian
\begin{equation}
\hat{H}_{L}=\hbar\left(\frac{\delta+A\cos(\omega_s t+\psi)}{2}\sigma_z+\frac{\Omega_{\vec{n}}}{2} \sigma_x\right). \label{H_L}
\end{equation}
As presentation in our recent experiment \cite{Chang_2020}, more than ten resolved Floquet sidebands are clearly observed. Both theory and experiment indicate that the initial phase $\psi$ is irrelevant to both Rabi oscillation and spectrum. 

\begin{figure}[t]
	\includegraphics[width=0.99\linewidth ]{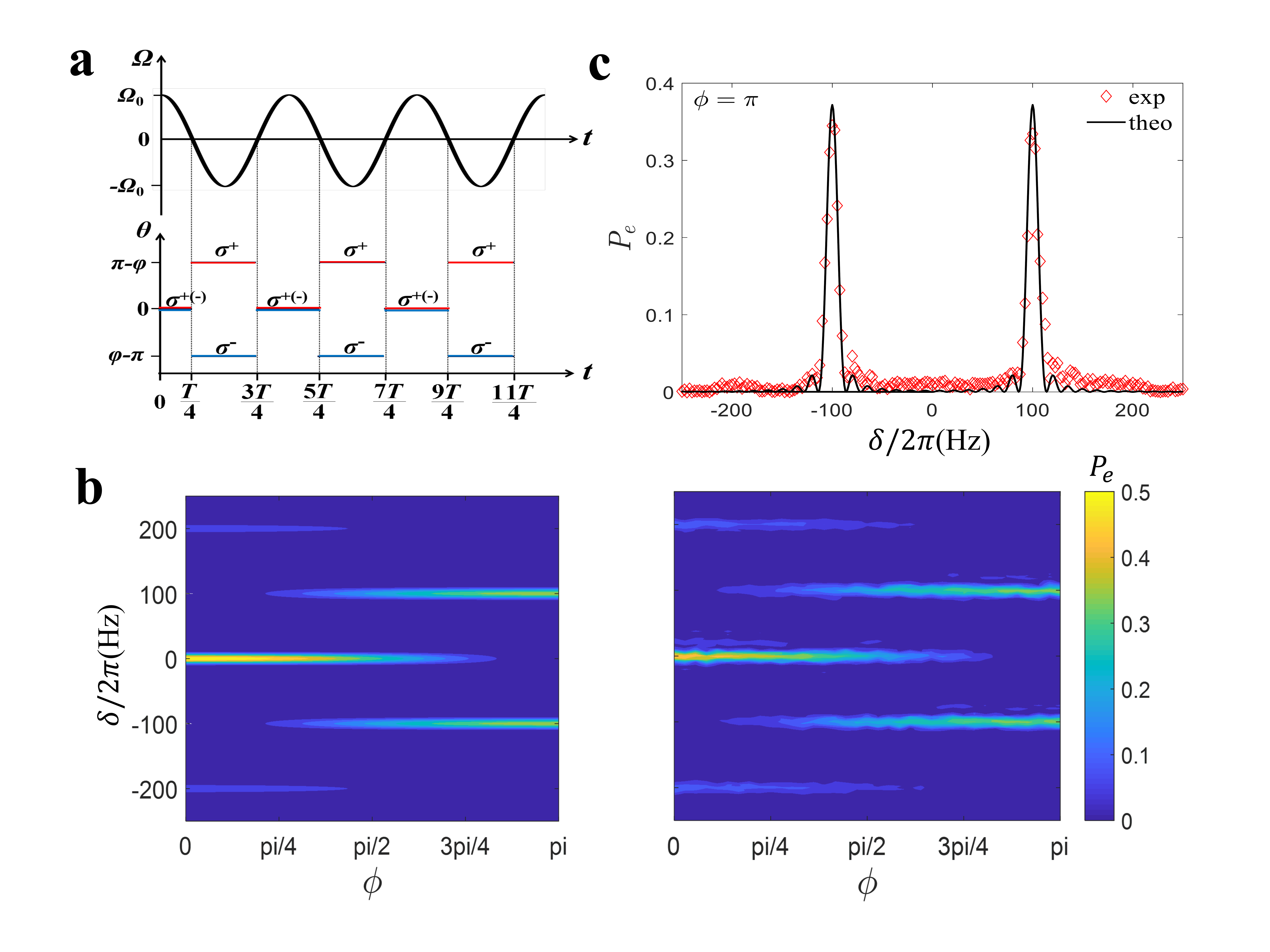}\hfill
	\vspace{-10px}
	\caption{\label{fig2} The transverse modulation. (a) The Rabi frequency is simultaneously under a sinusoidal waveform amplitude modulation and a square waveform phase modulation.  (b) is the Rabi spectrum of theoretical prediction (left) and experimental results (right) for phase $\phi$ changing from $0$ to $\pi$ at measuring time $t=70$ms. (c) is a comparison between experimental and theoretical Rabi spectrum with $\phi=\pi$. }
\end{figure}

\textit{Transverse modulation.--}To modulate the Rabi frequency, we implement the amplitude modulation on the clock laser. Because the Rabi frequency is proportional to the square root of clock laser's intensity $\Omega(t)\propto \sqrt{I(t)}$, we set the modulation function as S$_{\mathrm{PW}}(t)=\cos^2(\omega_s t)$. However, the additional $\pi$ phase is automatically inserted at the zero point $\Omega(t)=0$, and causes the discontinuity. To compensate the phase, as demonstrated in Fig.\ref{fig2}(a), we utilize the phase modulation following the square wave oscillating between 0 and $\phi$ \cite{SM}. Then the Hamiltonian after rotating wave approximation (RWA) can be written as
\begin{equation}
\hat{H}_T=\frac{\hbar\delta}{2}\sigma_z+\frac{\hbar \Omega_{\vec{n}}}{2}\cos\omega_st\left(\sigma_+e^{if(t)}+h.c.\right), \label{H_T}
\end{equation}
with the effective phase modulation function $f(t)$ as
\begin{equation}
f(t)=\left\{\begin{array}{lcl}
0&& {-T/4\le t-nT< \hspace{5px}T/4} \\
\pi-\phi && { \hspace{8px}T/4\le t-nT< 3T/4}
\end{array} \right.. \label{ft}
\end{equation}
{where $T=2\pi/\omega_s$ is the modulation period.} The Rabi frequency is under both amplitude and phase modulation as shown in Fig.\ref{fig2}(a). Because the experiment was run in the parameter region where the resolved sideband approximation $\omega_s \gg \Omega_{\vec{n}}$ holds, the lowest order effective Hamiltonian derived from Floquet-Magnus expansion for $n_t$th Floquet sidebands is expressed as
\begin{equation}
\hat{H}_{Te}^{n_t}=\frac{\hbar \delta_{n_t} }{2}\sigma_z+\left(\frac{\hbar \Omega_{\vec{n}}^{n_t}}{2}\sigma_++h.c.\right), \label{H_Te}
\end{equation}
in which $\delta_{n_t}=\delta-{n_t}\omega_s$ and the effective Rabi frequency is
\begin{equation}
\Omega_{\vec{n}}^{n_t}=\left\{\begin{array}{lcl}
\frac{\Omega_{\vec{n}}(1-\mathrm{e}^{-i\phi})}{4} && {{n_t}=\pm 1} \\
-\frac{\Omega_{\vec{n}}(1+\mathrm{e}^{i\phi})\mathrm{e}^{\frac{i {n_t}\pi}{2}}}{\pi(n_t^2-1)} && {{n_t}=2m} \\
0 && {\mathrm{others}}
\end{array} \right..
\end{equation}
Then, the probability of internal excited states is
\begin{equation}
P_e^{n_t}(\delta,t)=\sum_{\vec{n}}q(\vec{n})\left|\frac{\Omega_{\vec{n}}^{n_t}}{R^{n_t}_{\vec{n}}}\right|^2\sin^2\left(\frac{R^{n_t}_{\vec{n}}}{2}t\right),\label{Pe}
\end{equation}
where  $R^{n_t}_{\vec{n}} = \sqrt{|\Omega_{\vec{n}}^{n_t}|^2+\delta_{n_t}^2}$ with the Boltzmann distribution factor  $q(\vec{n})$ of external states $|\vec{n}\rangle$ \cite{Ye_2009}. 
In Fig.\ref{fig2}(b), the experimental Rabi spectrum at measuring time $t=70$ms matches well with the theoretical result. To remove the phase discontinuity of amplitude modulation, we set $\phi=\pi$ for total compensation. Demonstrated in Fig.\ref{fig2}(c), all the Floquet channels are totally suppressed down to weaker than noise, except for ${n_t}=\pm1$ Floquet channels with equal height \cite{SM}.

\textit{{Double modulation and interference.}--} Longitude and transverse modulations are demonstrated to be well fine-tuned, respectively. However, the modulation of one parameter should not affect another one for double modulation. Meanwhile, both modulations must be synchronized to make their relative phase $\psi$ finely tunable and stable \cite{SM}.
To achieve that, we use function generator to generate both modulation functions with the same frequency and chose it as low as $\omega_s/2\pi=100$ Hz.
 After the RWA, the system 
can be well described with explicit form of Eq. \ref{H}
\begin{equation}
\hat{H}_{D}=\hbar\left[\frac{\delta+A\cos(\omega_s t+\psi)}{2}\sigma_z+\frac{\Omega_{\vec{n}}}{2}\cos \left(\omega_s t \right) \sigma_x\right]. \label{H_D}
\end{equation}
Then, if we consider both modulations can produce a set of Floquet quasi-energy spectra with the same intervals, a question will immediately arise --- could relative phase make different Floquet channels interfere, so that the Rabi spectra changed?

\begin{figure}[t!]
	\includegraphics[width=0.99\linewidth ]{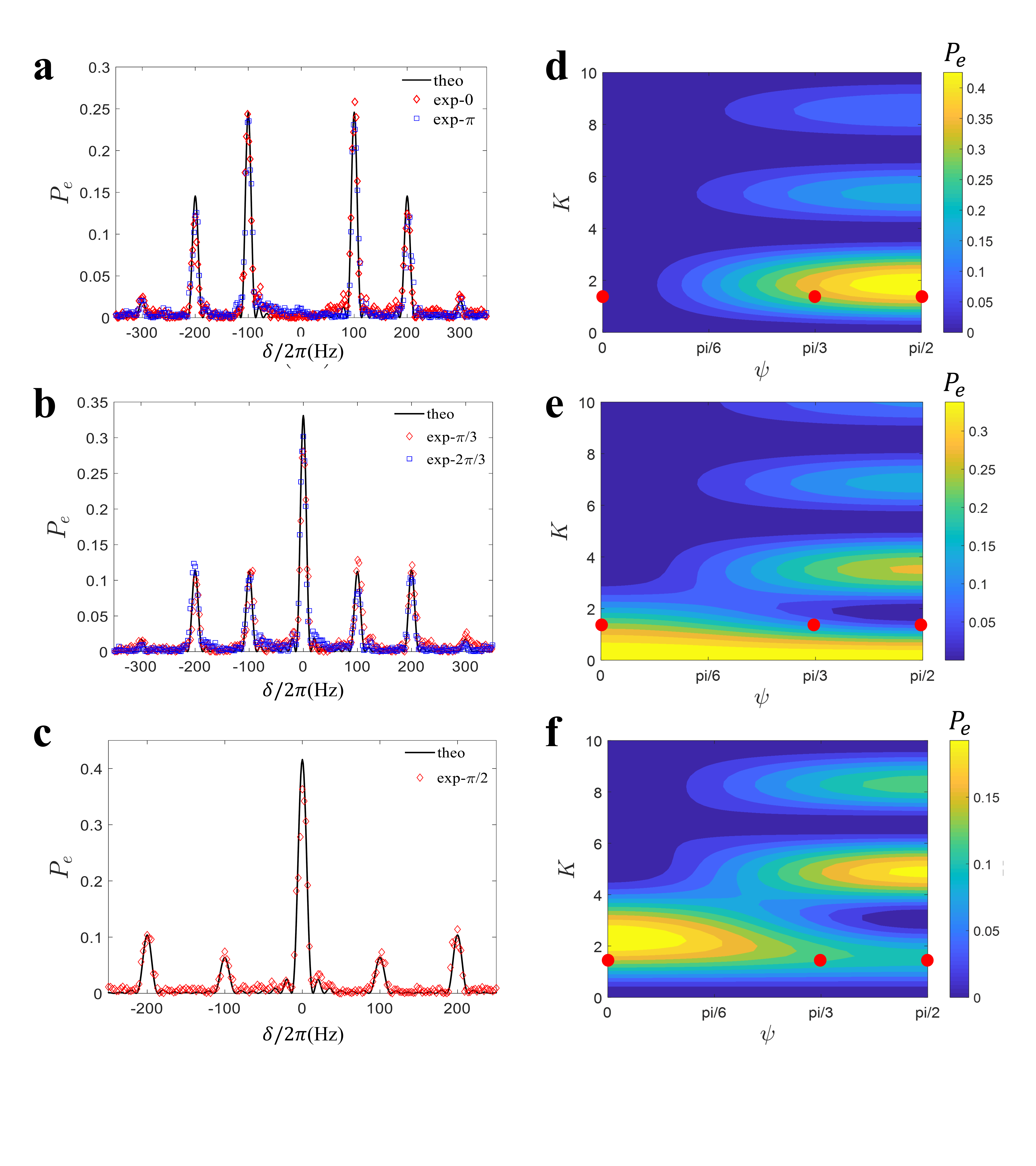}\hfill
	\vspace{-30px}
	\caption{\label{fig3} The Rabi spectrum of double modulation. (a-c) are the comparison between theoretical and experimental results for different $\psi$ at $K=1.38$. (d-f) are the peak's height of $0-2$th order Floquet sidebands at measuring time $t=70$ms calculated with Floquet theory, and the red dots mark the parameters chosen in (a-c).}
\end{figure}
In our experiment, we only vary the relative phase without changing other parameters. As shown in Fig.\ref{fig3} (a-c), the Rabi spectra strongly change according to $\psi$. There are two main features that can be clearly observed, (I) the different Floquet sidebands undergo different destructive or constructive interference processes, and (II) the spectrum at $\psi$ and $\pi-\psi$ are the same which reflects a hidden symmetry. In order to reveal its mechanism, utilizing the Jacobi-Anger relation and Floquet theory, we obtain the effective Floquet Hamiltonian similar as Eq. \ref{H_Te}, but with different effective Rabi frequency for $n_d$th Floquet side band
\begin{equation}
\Omega_{\vec{n}}^{n_d}=\frac{\mathrm{e}^{i{n_d}\psi}\Omega_{\vec{n}}}{2}\left(J_{{n_d}-1}[K]\mathrm{e}^{-i\psi}+J_{{n_d}+1}[K]\mathrm{e}^{i\psi}\right), \label{rabi_2ke}
\end{equation}
in which $K=A/\omega_s$ is the renormalized driving amplitude and {where $J_n[]$ is the $n$th order first kind Bessel function \cite{SM}}. Together with Eq.\ref{Pe} we could calculate the Rabi spectrum, and the Fig.\ref{fig3}(a-c) demonstrates the theoretical results are quite consistent with the experimental data.

The interference can be understood via expression of effective Rabi frequency Eq. (\ref{rabi_2ke}). The ${n_d}$th Floquet frequency has two terms which are related to ${n_d}\pm1$ Floquet modes of longitude modulation, respectively. As demonstrated in Fig.\ref{fig1} (b-c), in dressed atom picture, the longitude modulation provides lots of `Floquet photons' carrying $\psi$ phase. Meanwhile, if the transverse modulation is also on, the hopping from $|g\rangle$ to $|e\rangle$ states can have two channels $\omega_p^+$ and $\omega_p^-$. Because both modulations have the same driving frequency $\omega_s^x=\omega_s^z=\omega_s$, when the $\omega_p^++({n_d}-1)\omega_s=\omega_p^-+({n_d}+1)\omega_s$ is equal to the gap $\Delta$, the atoms can be excited via both channels. Considering the phase difference between both channels is $2\psi$, the atoms can interfere with each other and the period is $\pi$. From the theoretically calculated heights of $0-2$th Floquet sidebands peaks shown in Figs. \ref{fig3} (d-f), we can find that the interference of all Floquet sidebands could be observed by adjusting renormalized driving amplitude $K$. Then, because the Bessel functions are real, the strongest interference effect happens at $\psi$ equals to 0 and $\pi/2$. In addition, because the probability of excitation $P^{n_d}_e$ is irrelevant to the argument of Rabi frequency, the Rabi spectrum at $\psi$ and $\pi-\psi$ are the same. Furthermore, for different driving frequency, the Floquet spectrum will be nearly unchanged via rescaling detuning with unit of driving frequency (see supplementary material \cite{SM} for different $\omega_s$ at $K=1.1$).

\textit{{Double modulation and topology.}--}  The ${n_d}$th sideband effective Floquet Hamiltonian of doubly modulated OLC could also be taken as spin $1/2$ atom coupled to two magnetic fields $H_{\mathrm{EF}}^{n_d}=\frac{\hbar}2\vec{h}_{n_d}\cdot \vec{\sigma}$ with $\vec{h}_{n_d}=\vec{B}_{{n_d}-1}+\vec{B}_{{n_d}+1}$ and $\vec{B}_{n_d}=\frac{\Omega_{\vec{n}}J_{n_d}[K]}{2}\{\cos({n_d}\psi),-\sin({n_d}\psi),\frac{\delta-{n_d}\omega_s}{\Omega_{\vec{n}}J_{{n_d}}[K]}\}.$
Then we can consider ${h}_{{n_d}}^z=0$ where the peak of ${n_d}$th order Floquet sideband stands. Reminiscent of famous Su-Schrieffer-Heeger (SSH) model in momentum space  $H_{\mathrm{SSH}}=\vec{h}(k)\cdot\vec{\sigma}$ \cite{Su_1988,Kane_2010}, numbers of effective Hamiltonians $H_{\mathrm{EF}}^{n_d}$ can be provided for simulating various 1-D quantum topological phase after mapping the phase $\psi$ to quasi-momentum $k$. Similar to analyzing the endpoint of $\vec{h}(k)$ while letting the quasi-momentum $k$ walking around the Brillouin zone, we can also check different $\vec{h}_{{n_d}}(\psi)$ while varying $\psi$ from -$\pi$ to $\pi$. Meanwhile, the eigen-energies $E_{n_d}^{\pm}(\psi)=\pm\frac{\hbar|\vec{h}_{n_d}|}2$ can also be calculated for detecting the gap closed. Because the Rabi oscillation is only related to the modulus of effective Rabi frequency $|\vec{h}_{n_d}|$, we can experimentally measure the Rabi oscillation and extract the eigen-energies of different Floquet sidebands \cite{SM}.



\begin{figure}[t!]
	\includegraphics[width=0.99\linewidth ]{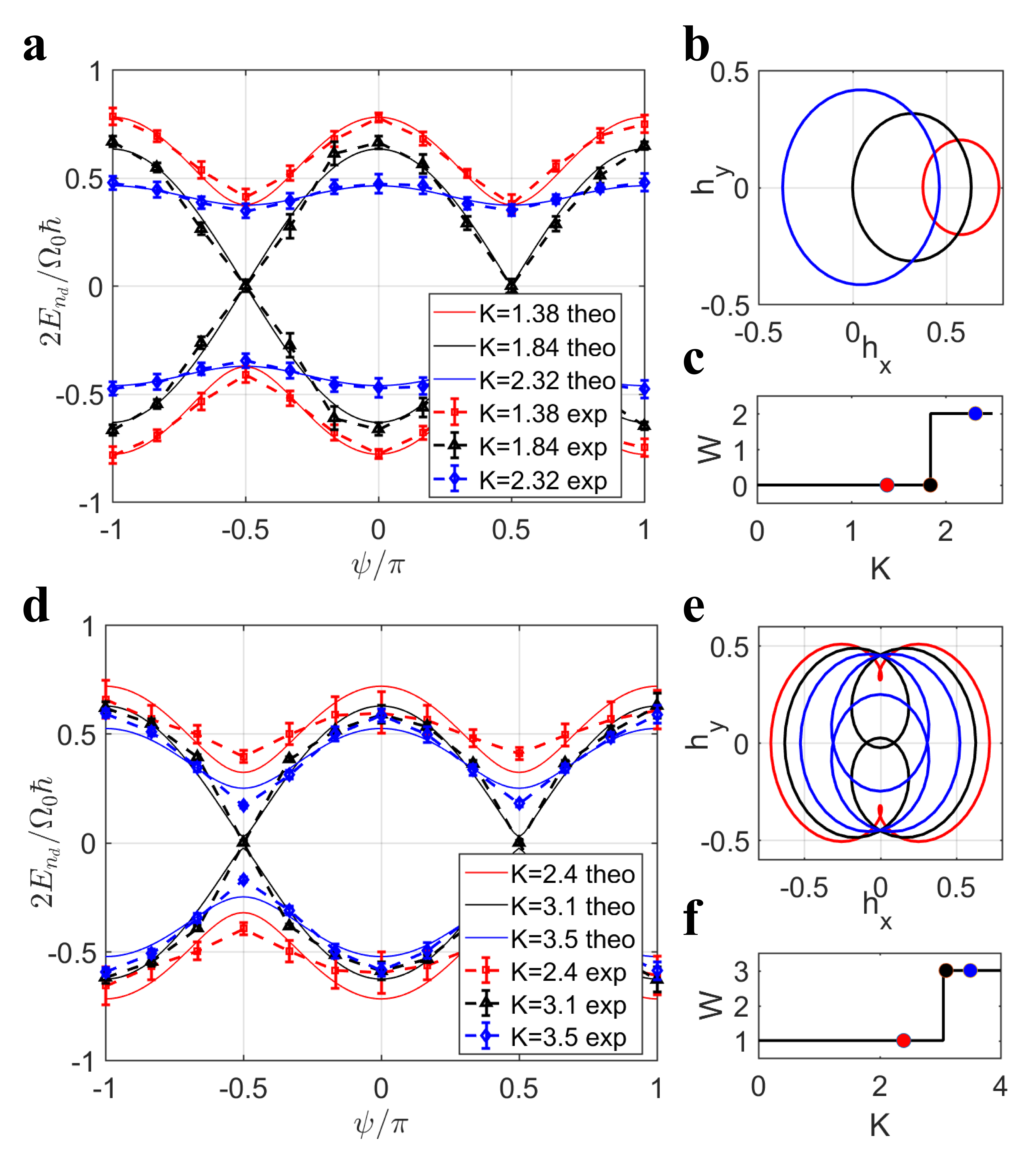}\hfill
	\caption{\label{fig4} The topology of Floquet effective Hamiltonians. (a) The eigen-energies in unit of bare Rabi frequency $\Omega_0$ of the 1st order Floquet sideband for different $K$ are obtained theoretically (solid line) and experimentally (dash line). The corresponding theoretical results of the `magnetic field' in $h_x-h_y$ plane and winding number are shown in (b) and (c) corresponding to same color in (a). The results of 2nd order Floquet sideband are shown in (d-f).} 
\end{figure}


For the zeroth Floquet sideband, the effective total magnetic field is $\vec{h}_0=\{0,-2J_1[K]\sin(\psi),0\}$ since $J_{-1}[K]=-J_1[K]$. Then because $\vec{h}_0$ only has non-zero $y$ component, the system can not have non-trivial topology. Turn to the first order Floquet sideband, in Figs. \ref{fig4}(a), {the experimental result of eigen-energies shows the energy gap closed at $K=1.84$. From the Rabi spectrum around it \cite{SM}, we obtain the experimental value with uncertainty $K_{c1}^E=1.84(8)$ which agrees well with the theoretical prediction $K_{c1}^T=1.8412$ }. When tuning the phase $\psi$ from -$\pi$ to $\pi$, the effective magnetic field can form a closed path. In Figs. \ref{fig4}(b), we can find the origin point moves from outside to inside by increasing $K$, and the touch point corresponds to the parameter at which the energies have the linear dispersion around the gap closed point $\psi=\pm\pi/2$. {The small deviations between theoretical and experimental results at low effective Rabi frequency is due to the dephasing of the Rabi oscillation\cite{Madej,Arey}.}
In order to quantitatively analyze the topology, we have to calculate the winding number $W=\frac{1}{2\pi i}\int^{\pi}_{-\pi} \frac{d\ln(h_{n_d}^x-i h_{n_d}^y)}{d\psi}d\psi$ which is a topological invariant. As shown in Fig.\ref{fig4}(c), winding number equal to zero and two corresponds to the origin point outside and inside the closed path, respectively. The linear dispersion 
is related to topological transition point from zero to two, and also closed path touching from negative $h_x$ direction. The topological transition of the second order Floquet sideband is shown in Figs. \ref{fig4}(d-f), and the winding number changes from one to three. The topological transition point {$K_{c2}^E=3.10(11)$} \cite{SM} is also related to linear dispersion of energy at $\psi=\pm\pi/2$. Differently, the closed path touches the origin point from both $\pm h_y$ directions.

These topological properties can be theoretically understood by analyzing two effective magnetic fields $\vec{B}_{{n_d}-1}$ and $\vec{B}_{{n_d}+1}$, of which the winding number is ${n_d}-1$ and ${n_d}+1$, respectively. Because the total magnetic field $\vec{h}_{n_d}$ is homotopic to one effective magnetic field which has larger magnitude, the winding number of the system is determined by the largest magnitude one. When the renormalized driving amplitude $K$ is small, the magnitude of $\vec{B}_{{n_d}-1}$ is larger than $\vec{B}_{{n_d}+1}$, so the winding number is ${n_d}-1$. Then, when $|\vec{B}_{{n_d}-1}|$ is equal to $|\vec{B}_{{n_d}+1}|$, we obtain the first critical point where winding number changes from ${n_d}-1$ to ${n_d}+1$. Considering $|\vec{B}_{{n_d}-1}|-|\vec{B}_{{n_d}+1}|$ alternative changing around zero, the winding number should oscillates between ${n_d}-1$ and ${n_d}+1$. 

\textit{{Conclusion and outlook.}--} We successfully realized double modulation in the OLC platform. By fine-tuning the relative phase between both modulations, we observed the clear interference effect induced by two types of `Floquet photons'. Meanwhile, we find they are strongly related to the 1-D topological insulator and also experimentally observed the energy gap open-close-open behavior. Beyond the SSH model, the high order effective Floquet Hamiltonians preserve higher winding numbers.  

{In the OLC platform, the coherence time could last at least more than 6s \cite{Ye_2017_2}}, so the topological phase transition and related physics of dynamics can be finely studied in the future work. In addition, the realization of double modulation can open several research paths. {The straightforward way is extending the simulation of topological insulator from 1-D to 2-D by tuning the compensated phase $\phi$. Meanwhile, more complex topological phase can be simulated by combining with other degree of freedoms \cite{Zhai_2020}.} On the other hand, choosing two driving frequencies incommensurate or opening the tunneling of nearest neighbor lattice sites \cite{Ye_2017} {may also bring in chaos or exotic many-body Hamiltonians.}


X.-F. Z. thank valuable discussions with Z.-J. Xiong and S.-X. Qin. This work is supported by the National Science Foundation of China (Grant No. 61775220),  the Key Research Project of Frontier Science of the Chinese Academy of Sciences (Grant No. QYZDB-SSW-JSC004), and the Strategic Priority Research Program of the Chinese Academy of Sciences (Grant Nos. XDB21030100 and XDB35010202). T. W. is supported by the Special Foundation for theoretical physics Research Program of China (No. $11647165$) and China Postdoctoral Science Foundation Funded Project (Project No.:$2020M673118$). X.-F. Z. acknowledges funding from the National Science Foundation of China under Grants No. 11804034, No. 11874094 and No.12047564, Fundamental Research Funds for the Central Universities Grant No. 2020CDJQY-Z003.


\begin{thebibliography}{100}
\bibitem{Eckardt_2017}
\href{https://journals.aps.org/rmp/abstract/10.1103/RevModPhys.89.011004}{A. Eckardt, \textit{Rev. Mod. Phys.} \textbf{89}, 011004 (2017).}

\bibitem{Kitamura_2019}
\href{https://www.annualreviews.org/doi/10.1146/annurev-conmatphys-031218-013423}{T. Oka, and S. Kitamura, \textit{Annu. Rev. Condens. Matter Phys.} \textbf{10}, 387-408 (2019).}

\bibitem{Kiefer_2019}
\href{https://journals.aps.org/prl/abstract/10.1103/PhysRevLett.123.213605}{P. Kiefer, F. Hakelberg, M. Wittemer, A. Berm\'{u}dez, D. Porras, U. Warring, and T. Schaetz, \textit{Phys. Rev. Lett.} \textbf{123}, 213605 (2019).}

\bibitem{Roushan_2016}
\href{https://www.nature.com/articles/nphys3930}{P. Roushan, \textit{ et al.}, \textit{Nat. Phys.} {\bf 13}, 146 (2017).}

\bibitem{Shu_2018}
\href{https://journals.aps.org/prl/abstract/10.1103/PhysRevLett.121.210501}{Zijun Shu, Yu Liu, Qingyun Cao, Pengcheng Yang, Shaoliang Zhang, Martin B. Plenio, Fedor Jelezko, and Jianming Cai, \textit{Phys. Rev. Lett.} \textbf{121}, 210501 (2018).}

\bibitem{Mukherjee_2018}
\href{https://journals.aps.org/prl/abstract/10.1103/PhysRevLett.121.075502}{Sebabrata Mukherjee, Marco Di Liberto, Patrik hberg, Robert R. Thomson, and Nathan Goldman, \textit{Phys. Rev. Lett.} \textbf{121}, 075502 (2018).}

\bibitem{Mei_2020}
\href{https://journals.aps.org/prl/abstract/10.1103/PhysRevLett.125.160503}{Feng Mei, Qihao Guo, Ya-Fei Yu, Liantuan Xiao, Shi-Liang Zhu, and Suotang Jia, \textit{Phys. Rev. Lett.} \textbf{125}, 160503 (2020).}

\bibitem{Wen_1989}
\href{https://journals.aps.org/prb/abstract/10.1103/PhysRevB.39.11413}{X. G. Wen, F. Wilczek, and A. Zee, \textit{Phys. Rev. B} {\bf 39}, 11413 (1989).}

\bibitem{Esslinger_2019}
\href{https://www.nature.com/articles/s41567-019-0615-4}{Frederik G\"{o}rg, Kilian Sandholzer, Joaquin Minguzzi, Remi Desbuquois, Michael Messer and Tilman Esslinger, Nat. Phys. \textbf{15}, 1161 (2019).}

\bibitem{Poletti_2014}
\href{https://journals.aps.org/prl/abstract/10.1103/PhysRevLett.113.183002}{ S. Greschner, L. Santos, and D. Poletti, \textit{Phys. Rev. Lett.} \textbf{ 113}, 183002(2014).}

\bibitem{Mintert_2019}
\href{https://journals.aps.org/pra/abstract/10.1103/PhysRevA.100.053610}{ H. Zhao, J. Knolle, and F. Mintert, \textit{Phys. Rev. A} \textbf{ 100}, 053610(2019).}

\bibitem{Knolle_2020}
\href{https://journals.aps.org/prl/abstract/10.1103/PhysRevLett.124.160604}{ H. Zhao, J. Vovrosh, F. Mintert, and J. Knolle, \textit{Phys. Rev. Lett.} \textbf{ 124}, 160604(2020).}

\bibitem{Yang_2016}
\href{https://iopscience.iop.org/article/10.1088/1367-2630/18/1/013005}{ Y. Zheng, and S. Yang, \textit{New Journal of Phys.} \textbf{ 18}, 013005(2016).}

\bibitem{Katori_2005}
\href{https://www.nature.com/articles/nature03541}{M. Takamoto, F.-L. Hong,  R. Higashi, and H. Katori, \textit{Nature} \textbf{435}, 321-324 (2005).}

\bibitem{Ye_2015}
\href{https://journals.aps.org/rmp/abstract/10.1103/RevModPhys.87.637}{A. D. Ludlow, M. M. Boyd, J. Ye, E. Peik, and P. O. Schmidt \textit{Rev. Mod. Phys.} \textbf{87}, 637 (2015).}

\bibitem{PoliNC_2013}
\href{https://www.sif.it/riviste/sif/ncr/econtents/2013/036/12/article/0}{N. Poli, C. W. Oates, P. Gill, and G. M. Tino, \textit{La Rivista del Nuovo Cimento} \textbf{36} 555 (2013).}

\bibitem{Katori_2003}
\href{https://journals.aps.org/prl/abstract/10.1103/PhysRevLett.91.173005}{H. Katori, M. Takamoto, V. G. Palchikov, and V. D. Ovsiannikov, \textit{Phys. Rev. Lett.} \textbf{91}, 173005 (2003).}

\bibitem{Westergaard_2011}
\href{https://journals.aps.org/prl/abstract/10.1103/PhysRevLett.106.210801}{P. G. Westergaard, J. Lodewyck, L. Lorini, A. Lecallier, E. A. Burt, M. Zawada, J. Millo, and P. Lemonde, \textit{Phys. Rev. Lett.} \textbf{106}, 210801 (2011)}

\bibitem{Ye_2018}
\href{https://journals.aps.org/prl/abstract/10.1103/PhysRevLett.120.103201}{G. Edward Marti, Ross B. Hutson, Akihisa Goban, Sara L. Campbell, Nicola Poli, and Jun Ye, \textit{Phys. Rev. Lett.} \textbf{120}, 103201 (2018)}

\bibitem{Ye_2020}
\href{https://www.nature.com/articles/s41567-020-0986-6}{Lindsay Sonderhouse, Christian Sanner, Ross B. Hutson, Akihisa Goban, Thomas Bilitewski, Lingfeng Yan, William R. Milner, Ana M. Rey, and Jun Ye, \textit{Nat. Phys.} \textbf{16} 1216 (2020)}

\bibitem{Safronova_2018} 
\href{https://journals.aps.org/rmp/abstract/10.1103/RevModPhys.90.025008}
{M. S. Safronova, D. Budker, D. DeMille, Derek F. Jackson Kimball, A. Derevianko, and C. W. Clark, \textit{Rev. Mod. Phys.} \textbf{90}, 025008 (2018).}

\bibitem{kolkowitz_2016}
\href{https://journals.aps.org/prd/abstract/10.1103/PhysRevD.94.124043}{S. Kolkowitz, I. Pikovski, N. Langellier, M. D. Lukin, R. L. Walsworth, and J. Ye, \textit{Phys. Rev. D} \textbf{94}, 124043 (2016).}

\bibitem{norcia_2017} 
\href{https://journals.aps.org/pra/abstract/10.1103/PhysRevA.96.042118}{M. A. Norcia, J. R. K. Cline, and J. K. Thompson, \textit{Phys. Rev. A} \textbf{96}, 042118 (2017).}

\bibitem{Ye_2017}
\href{https://www.nature.com/articles/nature20811}{S. Kolkowitz, L. Bromley, T. Bothwell, M. L. Wall, G. E. Marti, A. P. Koller, X. Zhang, A. M. Rey, and J. Ye, \textit{Nature} \textbf{542}, 66-70 (2017).}

\bibitem{Ye_2009}
\href{https://journals.aps.org/pra/abstract/10.1103/PhysRevA.80.052703}{S. Blatt, J. W. Thomsen, G. K. Campbell, A. D. Ludlow, M. D. Swallows, M. J. Martin, M. M. Boyd, and J. Ye, \textit{Phys. Rev. A} \textbf{80}, 052703 (2009).}

\bibitem{SM}
See Supplemental Material at ** for more details about the experimental realizations, Floquet methods, driving stability, Frequency shift, Eigen-energy measurement and uncertainty of $K_c$, which includes Refs. \cite{Ye_2009,pisa,R1_Campbell09,R1_Goban18,R1_Swallows12,R1_Martin13,Arey,Chang_2020}.

\bibitem{pisa}
\href{https://journals.aps.org/prl/abstract/10.1103/PhysRevLett.99.220403}{H. Lignier, C. Sias, D. Ciampini, Y. Singh, A. Zenesini, O. Morsch, and E. Arimondo, Phys. Rev. Lett. \textbf{99}, 220403 (2007).}

\bibitem{R1_Campbell09}
\href{https://science.sciencemag.org/content/324/5925/360.full}{G. K. Campbell \textit{et al.}, Science \textbf{324}, 360 (2009).}

\bibitem{R1_Goban18}
\href{https://www.nature.com/articles/s41586-018-0661-6}{A. Goban, \textit{et al.}, Nature \textbf{563}, 369 (2018).}

\bibitem{R1_Swallows12}
\href{https://ieeexplore.ieee.org/document/6174186}{M. D. Swallows, IEEE Trans.	Ultrason., Ferroelect., Freq. Contr. \textbf{59}, 416. (2012).}

\bibitem{R1_Martin13}
\href{https://science.sciencemag.org/content/341/6146/632.full}{M. J. Martin, M. Bishof, M. D. Swallows, X. Zhang, C. Benko, J. von Stecher, A. V. Gorshkov, A. M. Rey, and J. Ye. Science,	\textbf{341}, 632, (2013).}

\bibitem{Chang_2020}
\href{http://cpl.iphy.ac.cn/10.1088/0256-307X/38/7/073201}{Mo-Juan Yin, Tao Wang, Xiao-Tong Lu \textit{et al}  Chin. Phys. Lett. \textbf{38}, 073201 (2021)}



%
%


\bibitem{Su_1988}
\href{https://journals.aps.org/rmp/abstract/10.1103/RevModPhys.60.781}
{A. J. Heeger, S. Kivelson, J. R. Schrieffer, and W. P. Su, \textit{Rev. Mod. Phys.} {\bf 60}, 781 (1988).}

%


\bibitem{Kane_2010}
\href{https://journals.aps.org/rmp/abstract/10.1103/RevModPhys.82.3045}{ M. Z. Hasan, and C. L. Kane, \textit{Rev. Mod. Phys.} \textbf{82}, 3045(2010).}

\bibitem{Madej}
\href{https://dx.doi.org/10.1139/cjp-78-5-6-495}{L. Marmet and A.A. Madej, Can. J. Phys. \textbf{78}: 495.(2000).}

\bibitem{Arey}
\href{https://www.sciencedirect.com/science/article/abs/pii/S0003491613002546?via\%3Dihub}{A.M. Rey, A.V. Gorshkov, C.V. Kraus, M.J. Martin, M. Bishof, M.D. Swallows, X. Zhang, C. Benko, J. Ye, N.D. Lemke , A.D. Ludlow, Annals of Physics \textbf{340}, 311-351(2014).}

\bibitem{Ye_2017_2}
\href{https://science.sciencemag.org/content/358/6359/90.full}{S. L. Campbell, et.al., Science {\bf 358}, 90 (2017).}

\bibitem{Zhai_2020}
\href{https://www.nature.com/articles/s42254-020-0157-9}{Ren Zhang, Yanting Cheng, Peng Zhang, Hui Zhai,  Nature Reviews Physics,  \textbf{2}, 213 (2020).}


\end{thebibliography}
\end{document}